\documentstyle[preprint,prl,aps,epsfig]{revtex}

\begin{document}
\tightenlines

\begin{center}
\Large{\textbf{Thermally-induced expansion in the 8 GeV/c}}

\vspace{0.1cm} 
\Large{\textbf{$\pi^-$ + $^{197}$Au reaction.}}
\end{center}

\vspace{0.05cm}
\begin{center}
{T. Lefort$^1$}, 
{L. Beaulieu$^1$}, 
{A. Botvina$^{2}$}, 
{D. Durand $^{3}$}, 
{K. Kwiatkowski$^1$\footnote{Present address: Physics Division, 
Los Alamos National Laboratory, Los Alamos, NM 87545.}}, 
{W.-c. Hsi,$^1$\footnote{Present address: 7745 Lake Street, Morton Grove IL 
60053.}},
{L. Pienkowski$^4$}, 
{B. Back$^5$},
{H. Breuer$^6$}, 
{S. Gushue$^7$}, 
{R.G. Korteling$^{8}$}, 
{R. Laforest$^{9}$\footnote{Present address: Washington University Medical
School, 510 Kingshighway, St. Louis MO 63110.}}, 
{E. Martin$^{9}$}, 
{E. Ramakrishnan$^{9}$}, 
{L.P. Remsberg$^7$}, 
{D. Rowland$^{9}$},
{A. Ruangma$^{9}$}, 
{V.E. Viola$^1$}, 
{E. Winchester$^{9}$}, 
{S.J. Yennello$^{9}$}
\end{center}

\begin{center}
\it{$^1$Department of Chemistry and IUCF, Indiana University Bloomington, 
IN 47405 \\
$^{2}$ Institute for Nuclear Research, Russian Academy of Science, 117312 
Moscow, Russia.\\
$^{3}$ LPC de Caen, 6 Boulevard Marechal Juin, 14050 CAEN, France. \\
$^4$Heavy Ion Laboratory, Warsaw University, 02 097 Warsaw Poland.\\
$^5$Physics Division, Argonne National Laboratory, 9700 S. Cass Ave., 
Argonne, IL  60439.\\
$^6$Department of Physics, University of Maryland, College Park, MD 20742.\\
$^7$Chemistry Division, Brookhaven National Laboratory, Upton, NY 11973.\\
$^{8}$Department of Chemistry, Simon Fraser University, Burnaby, B.C., 
V5A 1S6 Canada.\\
$^{9}$Department of Chemistry and Cyclotron Laboratory, Texas A\&M  
University, College Station, TX 77843, USA.\\}
\end{center}

\begin{abstract}
Fragment kinetic energy spectra for reactions induced by 8.0 GeV/c 
$\rm{\pi^-}$ beams incident on a $\rm{^{197}}$Au target have been 
analyzed in order to deduce the possible existence and influence of 
thermal expansion. The average fragment kinetic energies are observed 
to increase systematically with fragment charge but are nearly 
independent of excitation energy. Comparison of the data with statistical 
multifragmentation models indicates the onset of extra collective thermal 
expansion near an excitation energy of E*/A $\rm{\approx}$ 5 MeV. However, 
this effect is weak relative to the radial expansion observed in 
heavy-ion-induced reactions, consistent with the interpretation that 
the latter expansion may be driven primarily by dynamical effects such as 
compression/decompression. 

\end{abstract}

\vspace{0.5cm}

\vspace{1cm}


The origin of the multifragmentation process \cite{Mor93}, and its link to 
a nuclear liquid-gas phase transition in finite systems \cite{Poc95}, is one 
of the most interesting and debated questions in the field of many-body 
nuclear dynamics. Is the fragmentation process thermally driven, initiated 
by an early compressional stage, or simply induced by mechanical or shape 
instabilities ? The observation of collective expansion energy at the 
end of the reaction may help to shed some light on the origin of the process.

The expansion of hot nuclear matter is usually attributed to either an
internal thermal pressure \cite{Des93} or the response to an initial 
compression produced at the beginning the reaction \cite{Rei97}. Two stages 
of the expansion can be schematically defined. The first drives the nucleus
up to the freezeout configuration, in competition with the restraining nuclear 
force. A possible second stage corresponds to an extra residual expansion 
energy (or radial flow) that exceeds the minimum required to reach 
freezeout. The collective expansion energy is 
proportional to the masses of the emitted particles.

The onset of extra expansion energy has been observed in heavy-ion collisions 
near 5-7 A MeV of available center-of-mass energy for fusion-like events 
($\rm{A_{tot} > 250}$) \cite{Riv99}. In a subsequent analysis,
Bougault et al \cite{Bou97} showed that a pure thermal extra expansion energy,
simulated with the Expanding-Evaporating Source model (EES) \cite{Fri90},
accounts for only a small part of the measured extra expansion energy.
On this basis and supported with BNV calculations, they attributed the extra
expansion energy observed in their data to an early compressional stage 
in the collision. Thus, one can link the 
multifragmentation energy threshold ($\rm{\approx 5~A MeV}$) to the onset 
of collective extra expansion energy initiated by a compressional phase
\cite{Riv99}. In this paper, we address the possible existence of 
thermally-induced extra expansion energy \cite{Des93} and its link to the 
thermal multifragmentation process. 

The existence of a thermally-induced extra collective expansion is 
an open question. The EOS collaboration found a large amount of collective 
expansion, up to 50 \% of the total available energy, in their study of 
$\rm{^{197}Au+~^{12}C}$ reaction at 1 A GeV \cite{Hau98}. On the other hand, 
the collective expansion observed in the spectator study 
of the ALADIN group is moderate \cite{Fri99}. In both cases, the 
excitation energy of the projectile should be mainly thermal, as for 
light-ion-induced collisions. Since the presence of collective expansion 
at high excitation energy may affect the isotope thermometer accuracy and 
hence the caloric curve shape \cite{Poc95}, it is important to determine the 
extent of collective expansion in these reactions.

The advantage of using light-ion-induced collisions stems from the nature 
of the deposited energy in the target nucleus: the contribution of 
compression, angular momentum and deformation is weak and the main part 
of the deposited energy is thermal \cite{Cug94,Wan96}. Previous studies 
\cite{Beau99} have shown that light projectiles can deposit excitation 
energies up to E*/A $\rm{\approx}$ 9 MeV in a gold target nucleus, well 
above the energy threshold for multifragmentation. Thus, light-ion-induced 
collisions offer a powerful tool for studying the relationship between 
multifragmentation and collective thermal expansion. 

In this letter we study fragment kinetic energy observables 
for the 8 GeV/c $\rm{\pi^-}$ + $\rm{^{197}}$Au reaction. The data
are compared with different statistical models: 
SIMON \cite{Dur92} and SMM (Statistical Multifragmenting Model) 
\cite{Bot90,Bon95}, as well as 
with data from heavy-ion reactions. The analysis is based on experiment 
E900a performed at the Brookhaven AGS accelerator with tagged beams of 8 GeV/c 
$\rm{\pi^-}$ + $\rm{^{197}Au}$ using the Indiana Silicon 
Sphere (ISiS), a 4$\rm{\pi}$ detector array with 162 
gas-ion-chamber/silicon/CsI telescopes \cite{Kwiat95b}. Further 
experimental details can be found in \cite{Lef99b}. 


In light-ion-induced collisions the emission spectra can be described 
with two components: an early pre-equilibrium emission stage that is 
forward-focused along the beam axis (mainly composed of energetic light 
charged particles) and isotropic emission from a slowly moving 
equilibrium-like residual source. Thermal-like charged particles 
are defined by the spectral shapes, from which an upper cutoff of 30 MeV 
for Z=1 and 9Z+40 MeV for heavier fragments is assigned \cite{Kwiat98}.
The pre-equilibrium-like particles emitted above the cutoff energy are 
removed. Then the charge, mass and excitation energy of the equilibrated 
residual source are determined via event-by-event reconstruction 
\cite{Beau99,Lef99b,Kwiat98}. The amount of pre-equilibrium emission 
increases with the excitation energy leading to a decrease of the 
equilibrated source average mass and charge, from $\rm{<Z>}$=76, 
$\rm{<A>}$=188 at E*/A=1 MeV to $\rm{<Z>}$=56, $\rm{<A>}$=138 at E*/A=9 MeV 
\cite{Beau99}. 

In \cite{Kun95,Rei97} the shape transition of the charge 
distribution, from a power-law behavior to an exponential-like pattern, 
has been observed in coincidence with an extra collective expansion energy. 
Since collective expansion may shorten the time for fragment formation, 
this transition has 
been interpreted as a sign of the expansion energy presence. For 
the 8 GeV/c $\rm{\pi^-}$ + $\rm{^{197}}$Au reactions, a power law is able 
to reproduce the charge distribution of the equilibrated component 
\cite{Beau99}, nonetheless, above E*/A=7 MeV, an exponential fit provides 
a better result. The exponential pattern is also observed with 
a pure statistical scenario \cite{Bon95} and is enhanced by secondary decay 
\cite{Wil97}. Therefore, it is necessary to investigate other observables
such as fragment kinetic energies in order to point out the existence of 
collective expansion. 

\begin{figure}
\centerline{\psfig{file=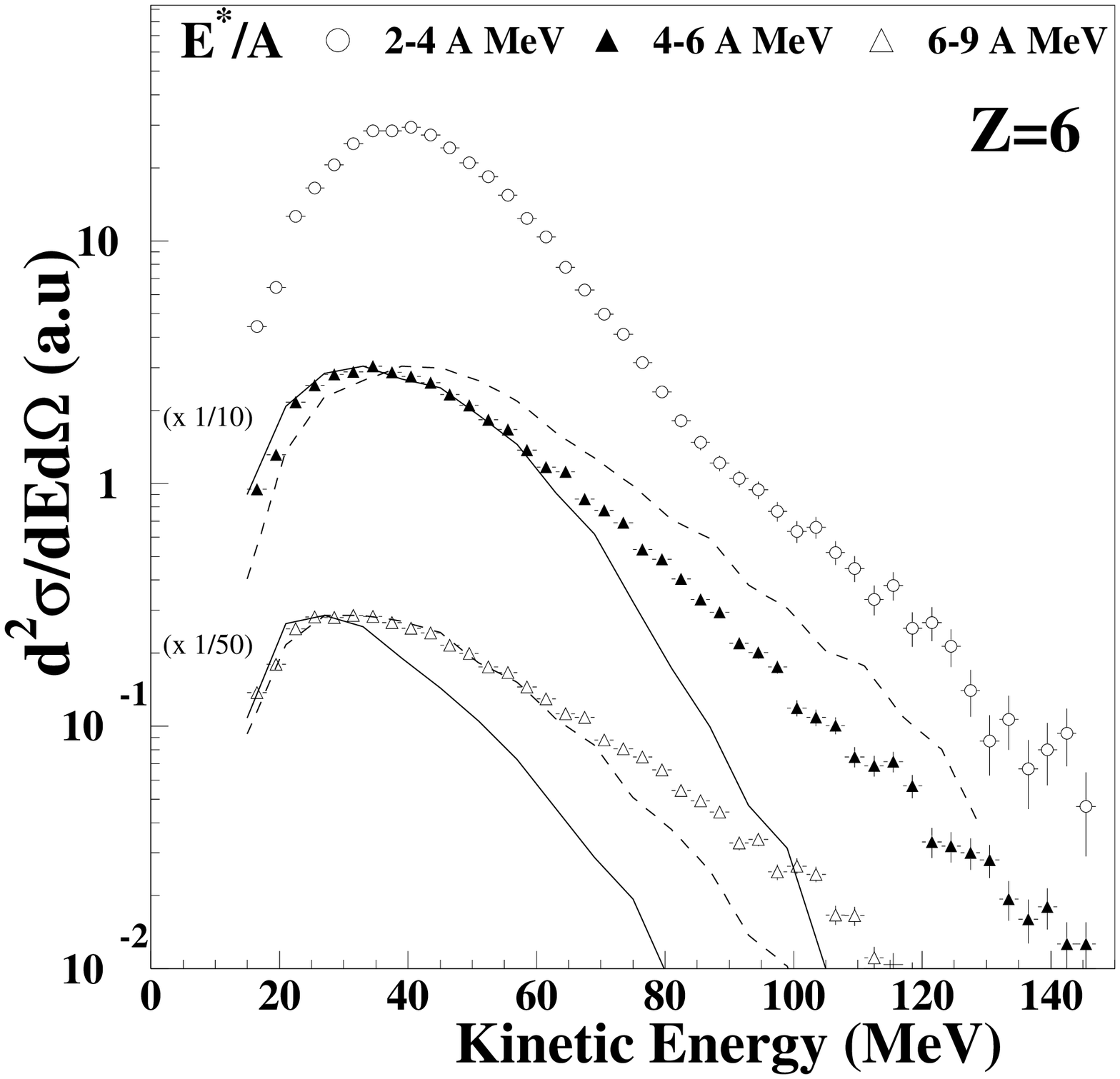,height=15cm,width=15cm}}
\caption{Angle-integrated kinetic energy spectra of carbon nuclei in the 
laboratory system for three bins of excitation energy, as indicated at the 
top of the figure. The lines correspond to SMM calculations ($\rm{\rho_0/3}$) 
with extra collective expansion energy, respectively, equal to 0 (solid line) 
and 0.5 A MeV (dashed line). For each bin in excitation energy, the simulated 
spectrum is normalized to the maximum of the experimental one.}
\label{Fig1} 
\end{figure}

In figure \ref{Fig1} the angle-integrated kinetic energy spectra for carbon 
nuclei (representative for all fragment spectra) are shown for three 
excitation energy bins. The energy of the Coulomb-like peak decreases  
with increasing excitation energy in agreement with the measured decrease 
of the source size (see above) and consistent with the onset of expanded 
nuclei. On the other hand the spectral slope increases with excitation 
energy. No evidence for a strong deformation of the spectra induced by a 
collective expansion is noticed, as was reported for heavy-ion-induced
reactions in \cite{Hsi94}.  

\begin{figure}
\centerline{\psfig{file=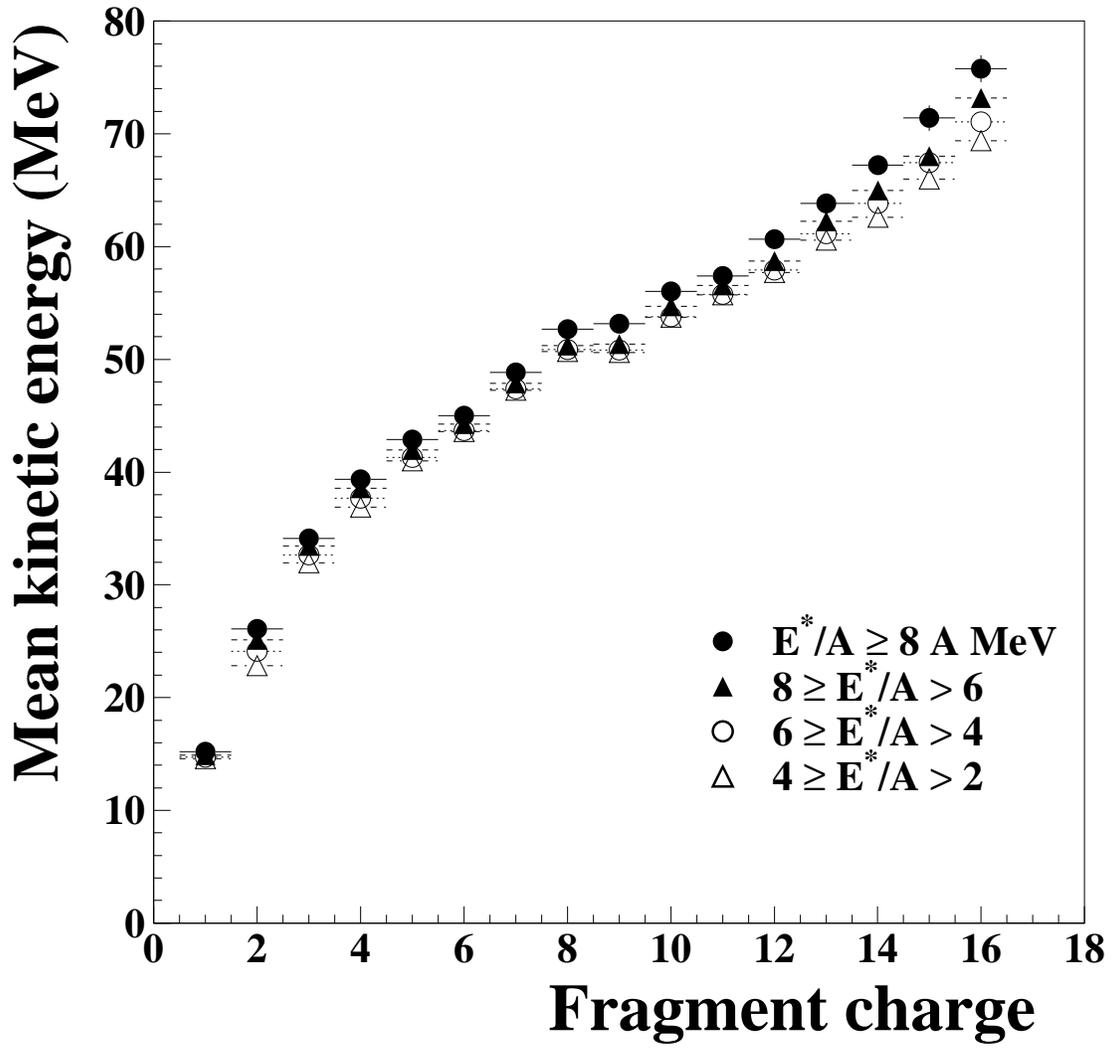,height=15cm,width=15cm}}
\caption{Fragment mean kinetic energy as a function of charge 
calculated in the source frame for 4 bins of excitation energy, as indicated 
on figure.}
\label{Fig2} 
\end{figure}

The mean kinetic energy of fragments as a function of their mass (charge) is 
also an indicator of the presence of collective expansion. One expects 
no dependence (flat behavior) for a pure thermal process, a slight increase 
due to Coulomb effects, and a steeper slope when an expansion energy 
is present. For a constant source size (charge) and density, the fragment 
mean kinetic energy is also expected to increase as a function of the 
excitation energy. In figure \ref{Fig2} the mean kinetic energy of  
fragments is plotted as a function of fragment charge for several 
excitation energy bins, transformed into the source frame. While 
the data are found to increase with the charge, little dependence on 
excitation energy is noticed. The constancy observed 
as a function of the excitation energy can be interpreted as a balance 
between the increase in thermal energy and the decrease in Coulomb repulsion 
of the emitting source (due to lower average source charge and possibly
density) with excitation energy, consistent with the evolution of the spectral 
shapes in figure \ref{Fig1}.  
Finally, it is worth mentioning that residues (if any), which have a 
lower average kinetic energy \cite{Bou97,Mar97,Bil99}, are not identified in 
ISiS due to threshold effects. 

\begin{figure}
\centerline{\psfig{file=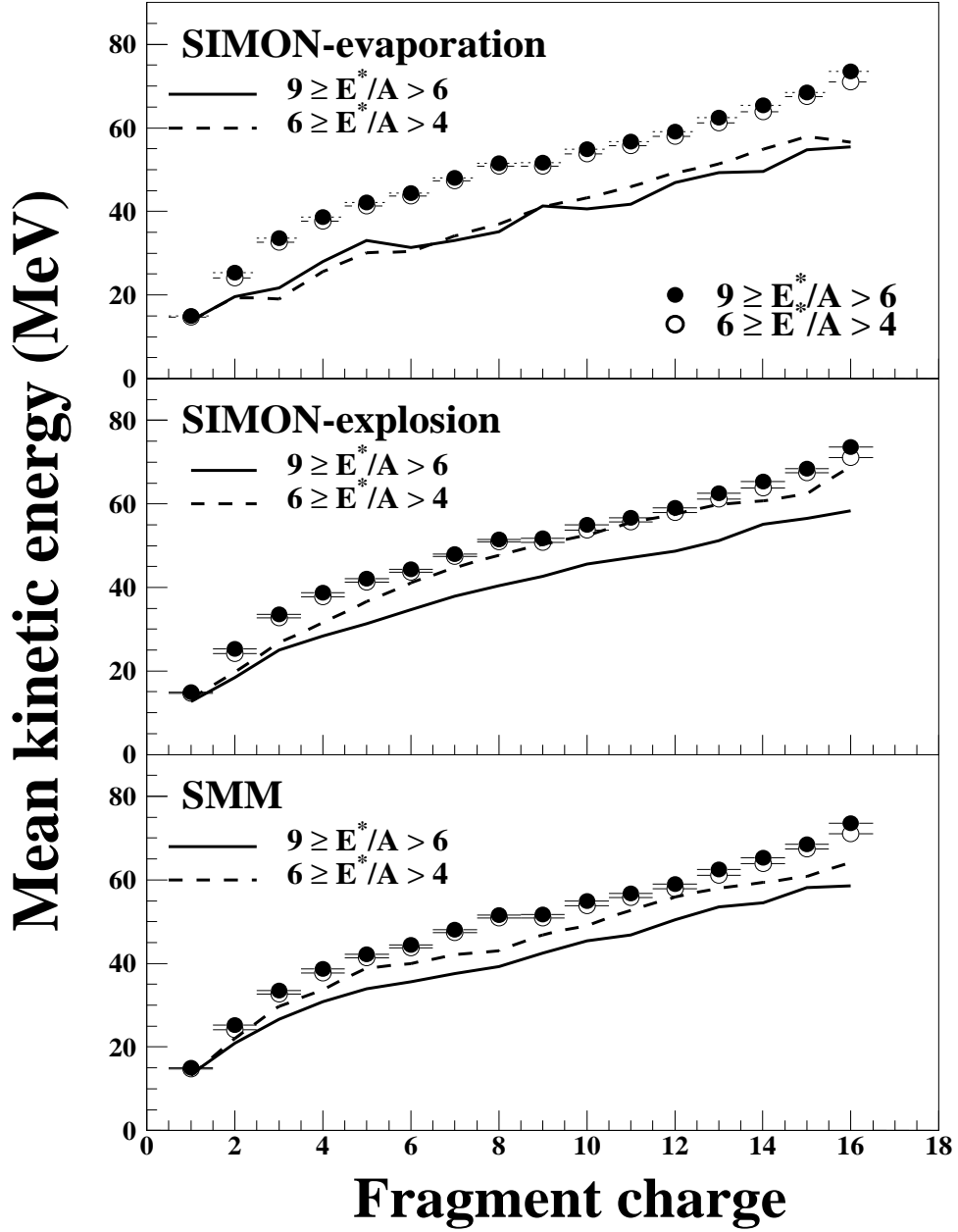,height=18cm,width=14cm}}
\caption{Comparison between experimental and simulated 
fragment mean kinetic energies calculated for two bins in excitation 
energy. In each panel, data are shown with open and solid circles 
and simulations with dashed and plain lines. The corresponding bins 
of excitation energy are indicated on the figure. SMM and SIMON-explosion 
calculations have been performed without additional expansion energy.}
\label{Fig3} 
\end{figure}

In this letter, we focus on excitation energies above E*/A=4 MeV, where one 
may expect to see evidence for collective expansion \cite{Riv99}. 
In figure \ref{Fig3}, the fragment mean kinetic energies are compared 
with predictions of SIMON-evaporation \cite{Dur92}, 
SIMON-explosion \cite{Dur92} and SMM \cite{Bot90,Bon95} simulations.  
The inputs of all the model simulations are identical, using the source 
charge, mass, velocity and excitation energy distributions reconstructed 
from data \cite{Beau99,Lef99b,Kwiat98}. 
Then the simulations are filtered to take account of the geometry 
of ISiS, the energy thresholds and the energy lost in the target. In 
addition, the simulated events have been sorted the same as in the 
experimental data and the excitation 
energy recalculated event-by-event. The results are found to be 
equal to the initial input within 10 \%, which gives an estimate of the  
confidence level of the comparison. The procedure was reduced to a 
single geometrical filter with SIMON-explosion. 

The procedure to extract the expansion energy is as follows. First we 
check to insure that the model reproduces the IMF 
multiplicity and charge distribution. If so, the thermal 
and Coulomb energies of fragments are calculated with the model 
and compared with the data. It may be necessary to add an extra collective 
energy (proportional to the mass of the emitted fragment) to reproduce 
the fragment mean kinetic energy. The extra collective energy corresponds
to the expansion energy.
     

Above E*/A=3 MeV, the evaporative model at normal density, 
SIMON-evaporation, 
underestimates the IMF (Z=3-16) multiplicity (by at least a factor two at 
E*/A=8 MeV), the mean kinetic energies and produces too few heavy IMFs.
This discrepancy confirms that a standard evaporative process is not able 
to reproduce the IMF multiplicities and the charge distributions at high 
excitation energy \cite{Wag99}, although this is still debated \cite{Gol96}. 
It is therefore not relevant to use this model to estimate an expansion 
energy.
 

Since the time-dependent evaporative model is unable to reproduce the data 
above E*/A=3 MeV, we then examined SMM and SIMON-explosion models, 
which assume a simultaneous break-up process. For both models the density 
is first set at one-third of normal density at the freeze-out stage. In our 
experimental event selection, we assume that the fast emission that takes 
place before the freeze-out is mainly removed by our cutoff energy defined 
above. Therefore, the experimental excitation energy determined within 
the energy cutoff is used as the source of thermal excitation energy at 
freeze-out for both models. 

In order to contrast between a picture in which the fragments are 
emitted cold and one where they are excited, we employ SIMON-explosion 
to investigate the former case and SMM for the latter. In the cold fragment
scenario, instead of feeding SIMON-explosion with a charge distribution of 
hot fragments that undergo secondary decay to produce the experimental charge 
distributions, the experimental charge distribution (cold fragments)
is used as input. In this context, the IMF multiplicty
and the charge distribution are in agreement with the data by definition.
Figure \ref{Fig3} shows that this model reproduces the data at 
E*/A=5 MeV but underpredicts the fragment kinetic energies at higher 
excitation energies.  

Finally, we used SMM in which it is assumed that thermal and collective 
expansion are unfolded \cite{Bon95} and only the thermal energy 
is used to generate partitions. The Z=6-16 multiplicity and 
charge distributions are well reproduced above E*/A=6 MeV. For both SMM 
and SIMON-explosion models, 
it is necessary to add about 0.5 MeV/A of collective energy for the 
E*/A=6-8 MeV bin in order to match the experimental mean kinetic energies. 
For SMM calculations at $\rm{\rho_0/3}$, good agreement with the kinetic 
energy spectra is obtained if an additional collective expansion energy 
is included, as shown by the lines in figure \ref{Fig1}. Except for the 
high kinetic energy tail, the carbon kinetic energy spectra are well 
reproduced at about 5 A MeV with no expansion energy (solid line) 
and with an added 0.5 A MeV (dashed line) for the E*/A=6-9 MeV bin.

\begin{figure}
\centerline{\psfig{file=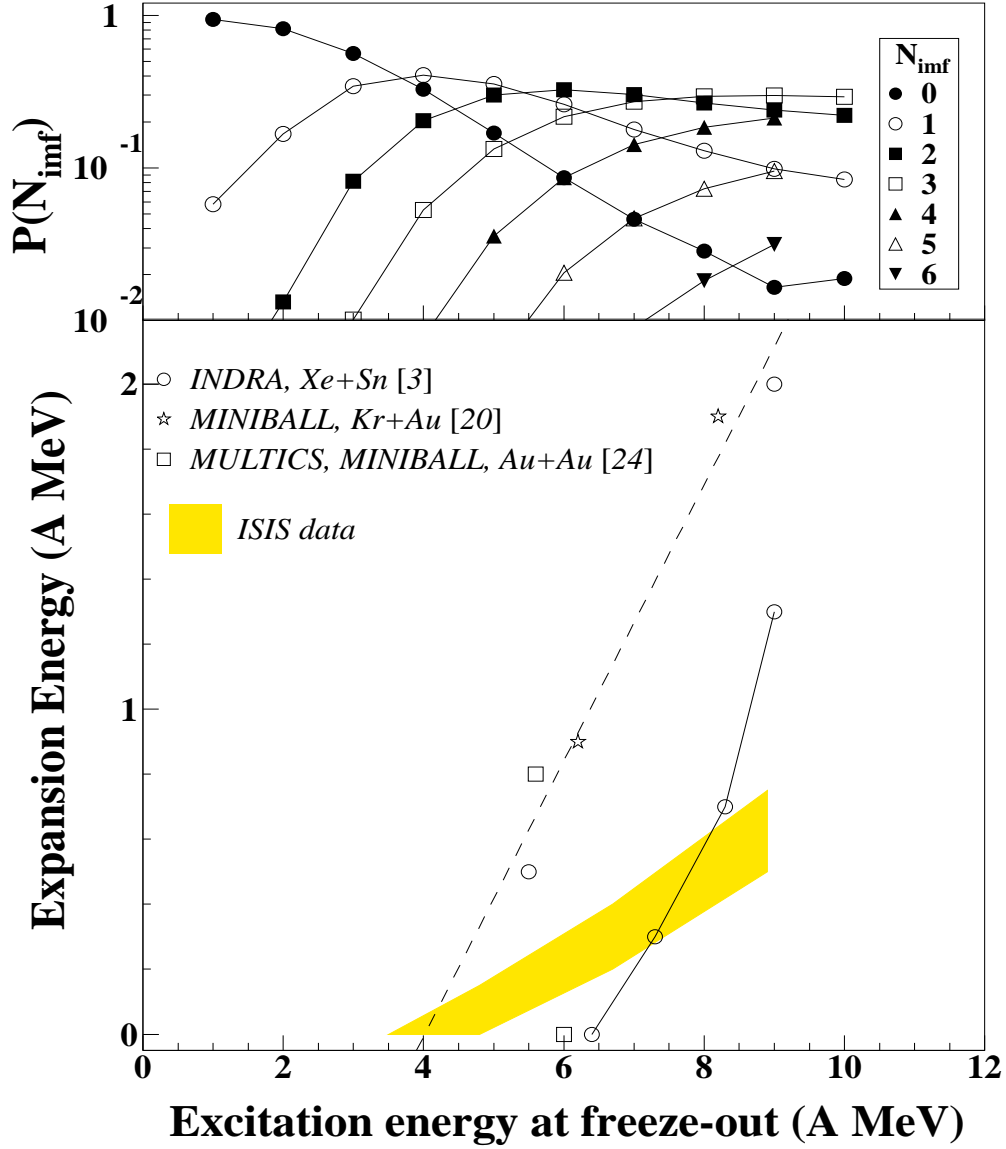,height=18cm,width=14cm}}
\caption{Upper panel: IMF emission probability as 
a function of the excitation energy. Lower panel: Comparison between 
central heavy-ion collisions and 8 GeV/c $\rm{\pi^-}$ + $\rm{^{197}}$Au 
reactions. The shaded area corresponds to the ISiS expansion energies 
extracted with SMM at $\rm{\rho_0/3}$ (upper limit) and $\rm{\rho_0/2}$ 
(lower limit). The dashed and plain lines set the boundaries of expansion 
energies extracted in central heavy-ion collisions with different assumptions 
regarding the source characteristics. See \protect\cite{Riv99,Wil97,Mad96} 
for more details.}
\label{Fig4} 
\end{figure}

As shown in figure \ref{Fig4} the amount of extra collective expansion energy 
is low and therefore highly dependent 
on the Coulomb energy in the simulations, i.e the source 
volume or density. In order to investigate the density dependence 
of the procedure used to extract the collective expansion, we have 
performed SMM calculations in which the density value is varied. 
The IMF multiplicity and the charge distribution predicted by SMM are 
in good agreement with such data for density values between $\rm{\rho_0/3}$ 
and $\rm{\rho_0/2}$. The two calculations, $\rm{\rho_0/3}$ 
and $\rm{\rho_0/2}$ correspond, respectively, to the upper and the lower
limit of the shaded zone in the lower panel of figure \ref{Fig4}. Even at 
a higher density, an additional collective energy is necessary to match 
the fragment mean kinetic energy. The onset of the extra collective 
energy occurs at about E*/A=5 MeV and  
increases with the excitation energy. This behavior is consistent 
with an increasing thermal pressure inside the nucleus as a function of the 
excitation energy. 

The upper panel of figure \ref{Fig4} shows the IMF emission probability 
as a function of the excitation energy. Below excitation energy of about 
E*/A=4 MeV the light charged particle emission is the prominent decay channel. 
At higher excitation energy, emission with one or more IMFs takes over. 
The onset of multiple IMF emission occurs in the same excitation 
energy range as that for the onset of the thermal expansion energy. The 
similarity underlines the possible link between expansion energy and 
multiple fragment emission probability.   


Finally, a comparison with heavy-ion collisions is also made in the lower 
panel of figure \ref{Fig4}. This comparison has been limited to systems with 
a well defined fusion source in order to avoid any problem of source 
separation present in the main part of the impact-parameter range. In 
addition, we only refer to studies performed in comparison with SMM 
\cite{Wil97,Mad96,Bou99}. The collective energy is expressed as a 
function of the freeze-out excitation energy (SMM input) instead of 
the available energy per nucleon used in \cite{Riv99}. In contrast with 
the small amount of thermal expansion energy found in ISiS collisions, 
in central heavy-ion collisions, the rise is much larger, as shown by the 
symbols and the two lines in figure \ref{Fig4} which correspond to 
different assumptions for extracting the collective expansion 
\cite{Wil97,Mad96,Bou99}. 
This behavior indicates that the collective expansion observed 
in central heavy-ion collisions cannot be explained with only a thermal 
component and supports the concept of an early compressional stage in 
central heavy-ion collisions that is not present in light-ion induced 
collisions. 


In conclusion, a study of the fragment kinetic energies has been performed 
for the 8 GeV/c $\pi^-$ + $^{197}$Au reactions. The sequential simulation 
at normal density, 
SIMON-evaporation, failed to reproduce the data above E*/A=4 MeV of 
excitation energy. For the two simultaneous models, SMM and SIMON-explosion, 
the fragment mean kinetic energies are well reproduced if an extra 
collective expansion energy is added at high excitation energy. Within the
context of the  
SMM calculation, the onset of this collective expansion energy 
takes place at about E*/A=5 MeV of excitation energy. The expansion energy 
increases slightly with the excitation energy, consistent with a 
thermally-induced expansion scenario. This observation, 
consistent with a soft explosion, also suggests that the nucleus is a 
dilute system at the break-up stage. Multiple IMF production takes place in 
the same excitation energy range, underlining the possible relationship 
between enhanced IMF emission and expanded nuclei. 
Nonetheless, the thermal expansion energy is weak and much lower than that 
observed in central heavy ion collisions within the same excitation 
energy range. Therefore, the main part of expansion energy observed in 
central heavy-ion collisions at intermediate energies must be related 
to a dynamical stage (initial compression ?) that does not exist in 
light-ion induced collisions.          

\textbf{Acknowledgements:} This work was supported by the U.S. 
Department of Energy and National Science Foundation, the National and 
Engineering Research Council of Canada, Grant No. P03B 048 15 of the 
Polish State Committee for Scientific Research, Indiana University Office 
of Research and the University Graduate School, Simon Fraser University 
and the Robert A. Welch Foundation.


\newpage

\newpage

\newpage

\newpage

\newpage

\end{document}